\documentstyle[psfig]{article}

\def\d{\mbox{d}}

\def\bra{\langle} 
\def\ket{\rangle} 
\def\a{\alpha} 
\def\b{\beta}
\def\dt{\delta}         
\def\D{\Delta}          
\def\L{\triangle}
\def\g{\gamma}          
\def\G{\Gamma} 
\def\e{\epsilon}
\def\ve{\varepsilon}

\def\lb{\lambda}

\def\m{\mu}
\def\n{\nu} 
\def\s{\sigma} 
 
\def\th{\theta}

\def\p{\partial} 
\def\f{\frac} 
\def\ov{\overline} 
\def\l{\left}
\def\r{\right} 
\def\Pl{\ell_{\sm{Pl}}}
\renewcommand{\t}[1]{\tilde{#1}}
\renewcommand{\^}[1]{\hat{#1}}
\newcommand{\sm}[1]{\mbox{\scriptsize #1}} 

\renewcommand{\@}[1]{\sqrt{#1}}
\def\be{\begin{eqnarray}}
\renewcommand{\le}[1]{\label{#1}
\end{eqnarray}} 
\def\ee{\end{eqnarray}}
\newcommand{\eq}[1]{(\ref{#1})} 
\def\nn{\nonumber\\} 

\newcommand{\rf}[1]{\cite{ref:#1}}
\newcommand{\rr}[1]{\bibitem{ref:#1}} 
 
\def\smqu{\ {\buildrel ?\over =}\ }
\def\ffract#1#2{\raise .35 em\hbox{$\scriptstyle#1$}\kern-.25em/
\kern-.2em\lower .22 em \hbox{$\scriptstyle#2$}}
\def\dts{\dt(\t\s-\t\s')} 
\def\ts{(\t\s-\t\s')}
\def\ftx{f(\t x-\t x')}
\def\fts{f(\t\s-\t\s')}

\def\dtx{\dt(\t x-\t x')}

\begin{document} 
\rm 
\large

\rightline{THU-98/23}
\rightline{gr-qc/9806028} 
\LARGE\vskip 1.8in\centerline{Planckian Scattering and Black Holes} 
\vskip 0.5in
\centerline{{\large Sebastian de Haro}{\large \footnote{\large
e-mail: {\tt haro@phys.uu.nl}}}} \vskip .3in\large 
\centerline{Spinoza Institute, Utrecht University} 
\centerline{Leuvenlaan 4, 3584 CE Utrecht}
\centerline{The Netherlands}
\vskip.1in
\centerline{and}
\vskip.1in

\centerline{Institute for Theoretical Physics} 
\centerline{Utrecht University} 
\centerline{Princetonplein 5, 3584 CC Utrecht}
\centerline{The Netherlands}

\rm
\large

\section*{}
Recently, 't Hooft's $S$-matrix for black hole evaporation, obtained from the 
gravitational
interactions between the in-falling particles and Hawking radiation, has been 
generalised
to include transverse effects. The action describing the collision turned out to 
be
a string theory action with an antisymmetric tensor background. In this article 
we show that the
model reproduces both the correct longitudinal and transverse dynamics, even
when one goes beyond the eikonal approximation or particles collide at 
nonvanishing incidence
angles. It also gives the correct momentum tranfer that takes place in the 
process. Including a
curvature on the horizon provides the action with an extra term, which can be 
interpreted as a
dilaton contribution. The amplitude of the scattering is seen to reproduce the 
Veneziano
amplitude in a certain limit, as in earlier work by 't Hooft. The theory 
resembles a
``holographic" field theory, in the sense that it only depends on the horizon 
degrees of
freedom, and the in- and out-Hilbert spaces are the same. The operators 
representing the
coordinates of in- and
out-going particles are non-commuting, and Heisenberg's uncertainty principle 
must be corrected
by a term proportional to the ratio of the ingoing momentum $p_{\sm{in}}$ to the 
impact
parameter $b$, $\D x \D p\geq\f{\hbar}{2}+{\cal O}(\Pl^2\,p_{\sm{in}}/b)$.
Reducing to 2+1 dimensions, we find that the coordinates satisfy an SO(2,1) 
algebra.
\section*{Introduction}

Over the last years, the relation between black holes and strings has been 
stressed both from the string theory side \rf{string} and from 
gravity \rf{gnato}. This has led to the hypothesis \rf{gdimred} that 
two-dimensional closed
surfaces (in four dimensions), and in particular black hole horizons, may carry 
all the
information that is hidden inside them. One of the concrete attempts to realise 
this idea (see
also \rf{malda} for more recent developments) is 
't Hooft's $S$-matrix description of black holes \rf{g9607}. He calculated the 
microscopical $S$-matrix for scattering between in-falling particles and 
out-coming 
radiation in a black hole background. The essential ingredient was the 
description of the dynamics of the horizon, that is deformed by the presence of 
the particles, thus modifying the trajectories of out-coming particles. It 
turned out that the action which appears in this $S$-matrix contains the 
Nambu-Goto action of the membrane (without time) or Euclidean string. In 
reference \rf{s}, the $S$-matrix was covariantly generalised.
Nevertheless, as stressed in that paper, since the 
starting point was the $S$-matrix as calculated in Minkowski space, an 
eventual curvature in the transverse direction was not taken into account. 
Furthermore, the original $S$-matrix had been calculated in the limit where the 
transverse momenta can be neglected, which corresponds to a large impact 
parameter, so it
was not obvious that the new expression would include 
the effect of these momenta. In this paper we study both problems in more 
detail, describing
highly energetic particles by means of 2-dimensional field theory.

In the first section we review the classical analysis of geodesics in the 
gravitational
field of a massless particle, and the Dray-'t Hooft treatment of the transverse 
curvature. More
details can be found in the appendices. We use these results in section 2 to 
calculate the
momentum transfer during the collision. Our approach is perturbative in the 
ratio of the
in-going momentum to the impact parameter times Planck's length squared. Then in 
section 3 we
check that the covariant equations of motion reproduce the trajectories of this 
scattering
process, both the longitudinal and the transverse ones. In section 4 we show how 
to include a 
curvature in the $S$-matrix. We find that, in string theory language, this 
amounts to a dilaton
term in the action, with a specific expression for the dilaton in terms of the 
positions of the
particles. There is thus a strong similarity between this model for Planckian 
scattering and
a $3+1$-dimensional string theory with a graviton, an antisymmetric tensor and a 
dilaton.
Working out the scattering amplitude, in the limit of vanishing transverse 
momentum one gets the
Veneziano formula.

In section 5 we study the quantum mechanics of the model. We find that Hilbert 
space is severely
reduced due to the identification between momentum and position operators. 
Actually, the
Hilbert space of the in-going particles is identified with that of the 
out-coming ones.
Their trajectories must be described by non-commuting coordinates. We also get
a noncommutative algebra for the in- and out- momentum operators, in particular 
for the
transverse ones, which satisfies the Jacobi identity. We find
evidence that Heisenberg's uncertainty principle has to be corrected with a term
proportional to Newton's constant. This is a consequence of a certain ``merging" 
of the different
coordinates due to the gravitational interaction. Indeed, during the collision 
there is a
momentum transfer which is reproduced by this new term. Although our results are 
not totally
conclusive, since the nonlocality of certain expressions poses constraints on 
the validity of
the model and leads to several technical problems, the proposal does give 
correct results within
the used approximation. We verify that the canonical momentum operator contains 
the information
about the recoil of the particles due to their mutual interaction. Another check 
of the
model is performed in the last section, where we compactify one dimension and 
compare with the
$2+1$-dimensional case, where one can overcome the mentioned problems. We find 
exact agreement.
In the lower dimensional case, the position operators obey an SO(2,1) algebra. 
Also in this case,
the Heisenberg uncertainty principle has to be modified.
\section{Classical scattering at Planckian energies}
The gravitational back-reaction of highly energetic particles on the
metric is the starting point of the $S$-matrix Ansatz. The underlying philosophy 
is that the
back-reaction of the in-falling particles modifies the out-coming trajectories. 
This
modification would allow one to make statements about the quantum state of the 
black hole.
The same holds for the back-reaction of out-coming Hawking radiation, which 
modifies the location
of the horizon. It was shown by Dray and 't Hooft \rf{gnp85} that the effect of 
a
particle on the metric is to shift the geodesics with a translation factor 
that is proportional to the momentum of the incoming particle, as we can see in 
Fig. 1.
This phenomenon is known as shockwave. In this section we review some known 
facts about
shockwaves, which we use in later sections.
\begin{center}
        \begin{figure}[h] \hspace{4cm}
        \psfig{figure=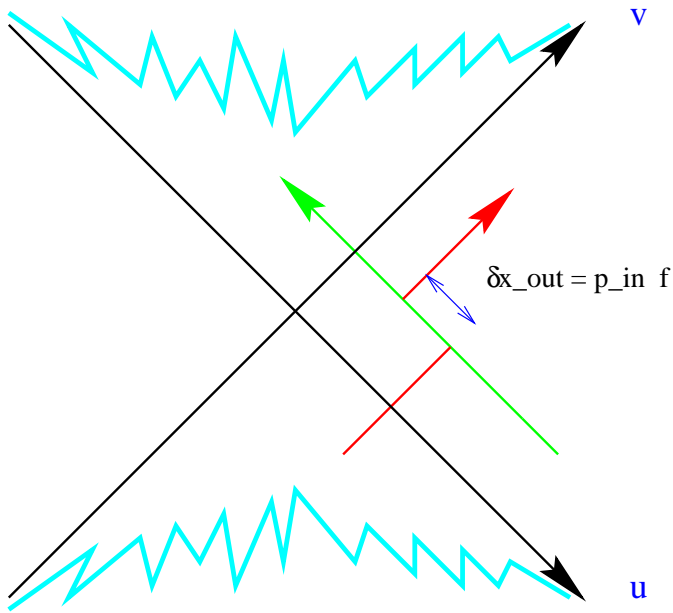}
        \caption{Effect of the shockwave in the lightcone directions}
        \end{figure}
\end{center}

The strength of the gravitational shift is given by\be
\delta x_{\sm{out}}= 8\pi G\,p_{\sm{in}}\,f(\t x),
\le{1}
where $p_{\sm{out}}$ and $x_{\sm{in}}$ are, respectively, the momentum and 
positions of the
out- and in-going particles in the longitudinal direction, and $\t x\equiv(x,y)$ 
are the
directions transverse to the motion of the particles, that are assumed to 
collide almost
frontally. $f$, the shift function, is the propagator of the two-dimensional 
surface
transverse to the trajectory of the incoming particle. In flat Minkowski space,
\be
\d s^2=2\d u\d v+\d x^2+\d y^2,
\le{1b}
it is given by
\be
f=\f{1}{4\pi}\,\log\t x^2.
\le{2}
For more details, we refer the reader to \rf{gnp85}.

Furthermore, there is a deflection of the trajectories in the transverse plane 
(see Fig. 2), described by
\be
\cot\a+\cot\b=8\pi G\,p_v\p_yf(y),
\le{3}
where we have chosen our axes so that the deflection is on the $y-v$-plane. So
we see that the function $f$ fully describes the geometry, since it gives 
us 
both effects. If we define light-cone coordinates $u=-t+z$ and $v=t+z$, then 
before the
collision the in-going particle travels along the trajectory $t=-z$, and the 
out-coming 
particle along $t=z$. The whole effect can then be 
captured in the metric
\be
\d s^2=2\d v\l(\d u-{\sf f}_v(\t x)\,\dt(v)\d 
v\r)+\d x^2+\d y^2,
\le{4}
where ${\sf f}_v(\t x)\equiv-\f{1}{T}\int\d^2\t x'P_v(\t x')\,f(\t x-\t x')$, 
with 
$T\equiv\f{1}{8\pi G}$. ${\sf f}$ is now the shift function. This is a 
straightforward generalisation
for the case that the total momentum is not concentrated at one point, but is a
distribution over the shockwave. This allows us to describe an arbitrary amount 
of
ingoing particles all sitting on the plane $v=0$\footnote{When going to quantum 
mechanics,
this condition will be relaxed.}, travelling with total momentum $P^u=P_v$. The 
{\it i}th
out-coming particle has an initial momentum $p^v_i=p_u^i$, and the in-going 
momentum distribution
$P_v(\t x)$ is typically equal to
\be
P_v(\t x)=\sum_{i=1}^Np_v^i\dt(\t x-\t x^i),
\le{4b}
if there are $N$ particles with transverse positions $\t x^i$ on the plane of 
the 
shockwave. All particles satisfy the mass-shell relation $p_up_v=0$.

One can reproduce these two effects by solving the Euler-Lagrange equations for
this metric. The mathematical subtleties of dealing with a metric including
distributions have been analysed in \rf{steinbauer}. Here, however, we will
suffice with a more heuristic treatment, which is good enough in the simple 
cases we will be considering. The first of the equations of motion in this 
metric gives
\be
\ddot{v}=0,
\le{5}
where the dot denotes the derivative with respect to the affine parameter $\lb$ 
along the geodesic. This allows us to use $v$ as a time coordinate. The other 
equations then
amount to the following solutions:
\be
u(v)&=&u(0)-\f{1}{2T}\,\mbox{sgn}\,(v)\int\d^2\t x'\,P_v(\t x')\l(f(\t x_0-\t 
x') 
+v\f{\p x^i}{\p v}(0)\,\p_if(\t x_0-\t x')\r)\nn
x^i(v)&=&x^i(0)+p^i_0v+\f{1}{2T}\,v\,\mbox{sgn}\,(v)\,\int\d^2\t 
x'\,P_v(\t x')\,\p_if(\t x_0-\t x'),
\le{7}
where we defined $\t x_0\equiv \t x(0)$ and used the identity 
$f(v)\,\dt(v)=f(0)\,\dt(v)$. As a
boundary condition, we have chosen that the initial momentum in the 
$u$-direction is zero,
and in the transverse $i$-direction\footnote{The latter will often be taken 
equal to zero.} it
is $p^i$.

If we now concentrate on the $y-v$ plane, and filling in equation \eq{4b}, 
differentiating
\eq{7} yields
\be
\f{\p y}{\p v}=\f{1}{2T}\,\mbox{sgn}\,(v)\, \int\d^2\t x'P_v(\t x')\,\p_yf(\t 
x_{\sm{0}}-\t x'),
\le{l8}
which reduces to the result outlined in appendix A if there is just one particle 
going in. The latter is based on the direct 
calculation of geodesics in the linearised Schwarzschild field of a light 
particle, after which the massless limit is taken, together with the limit that 
the particle travels at the speed of light.

In the following we will linearise our expressions in the expansion parameter 
$\ve$, defined as
$\ve\equiv G\,p_{\sm{in}}b$, where $p_{\sm{in}}$ is the in-going momentum and 
$b$ the impact
parameter, given by the transverse separation between the colliding particles. 
Notice that,
since $f$ is logarithmic in the transverse distance, $\p_if\sim\f{1}{b}$.

The first of \eq{7} gives us the shift \eq{1} in the longitudinal coordinate $u$ 
as a consequence of the ingoing particle plus a correction that ---if the 
ingoing particles 
have no momentum along the transverse directions---, as it is ${\cal O}(\ve^2)$,
can be neglected . The second of \eq{7} represents the kink in the trajectory of 
the out-coming
particle, equation \eq{3}.

\begin{center}
        \begin{figure}[h] \hspace{4cm}
        \psfig{figure=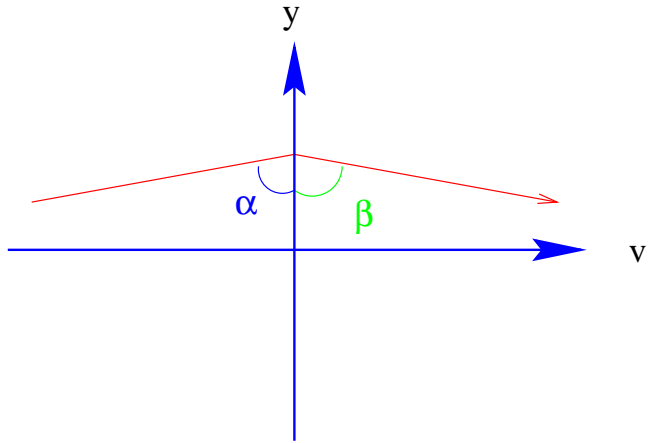}
        \caption{Effect of the shockwave in the longitudinal direction}
        \end{figure}
\end{center}

This method for calculating geodesics by a shift in the metric, which is 
Penrose's
scissors-and-paste method \rf{penrose}, applies to more 
general space-times. Indeed, starting with a metric of the form
\be
\d s^2=2A(u,v)\,\d u\d v+g(u,v)\,h_{ij}(\t x)\d x^i\d x^j,
\le{8}
which is a vacuum solution of Einstein's equations, Dray and 't Hooft showed 
that one obtains a solution describing a photon sitting at $v=0$, $(x,y)=0$, by
shifting $u$ by a factor of $p_v f$, if at $v=0$ the following conditions
are fulfilled\footnote{We have interchanged the $u$ and $v$ coordinates with 
respect to the Dray-'t Hooft's notation, so that the photon goes {\it into} 
the black hole if we consider Kruskal coordinates. We also have a slightly 
different definition for $f$: $\@{\det\,g_{\m\n}}\,f=-T\,G$, where $G$ is the 
Green's function used by Dray and 't Hooft.}:
\be
A_{,\,u}&=&0=g_{,\,u},\nn
\f{A}{g}\L f-\f{g_{,\,uv}}{g}f&=&\f{A^2}{\@{\det\,g_{\m\n}}}\,\dt^{(2)}(\t x-\t 
x_{\sm{0}}),
\le{9}
with the Laplacian $\L=\f{1}{\@{h}}\,\p_i\@{h}h^{ij}\p_j$. In the Minkowski
case, the solution of the second of \eq{9} is given by \eq{2}.
\section{Beyond the eikonal approximation: momentum transfer}

Until now we have discussed how the trajectories of out-coming particles are 
modified by the shockwaves of the ingoing particles. Next we will consider the 
momentum transfer,
which is related to the ingoing momentum, and we will need later to compare with 
the results
coming from the $S$-matrix.

From Fig. 2, we learn that
\be
\tan\g=\f{p_y}{p_u},
\le{-1}
where the angle $\g$ is defined by $\g=\pi-\a-\b$, and $\a$ and $\b$ are defined 
as in the
figure. $p_y$ and $p_u$ are the momentum 
of the out-coming particle in the $y$ and $v$-directions, respectively, {\it 
after} it passes the shock wave. These quantities are different from the momenta 
before the interaction, which we denote by $p_\m^0$. We do not explicitly write 
the superscripts in or out, since it should be clear from the context whether 
the momentum refers to the in-going or out-coming particle\footnote{In this 
section we assume 
there is only one particle coming in.}.

We now take the initial transverse momentum to be zero, $p_y^0=0$. This means 
that $\a=\pi/2$ and hence
\be
\cot\a+\cot\b=\tan\g=\f{1}{T}\,p_v\,\p_yf(y_0).
\le{-3}
One can easily calculate that the exchange of momentum in the $v$-direction, to 
first order in
$\ve$, is equal to zero, so that $p_u\simeq p_u^0$. Therefore, we have
\be
p_y=\f{1}{T}\,p_up_v\p_yf.
\le{-4}
Since $p_y\sim\f{\p y}{\p v}$ and the particle is travelling at the speed of 
light, 
this can also be directly deduced from the second of \eq{l8} (see section 3).

If the initial transverse momentum is nonzero, differentiating \eq{7} once 
yields
\be
\f{\p u}{\p v}=-\f{1}{2T}\,\mbox{sgn}\,(v)\f{\p x^i}{\p v}\,p_v\p_if(\t x_0),
\le{-5}
so for the out-coming particle we have
\be
p_v=-\f{1}{T}\,p^i_0p_v\,\p_if(\t x_0).
\le{-6}
This can also be obtained from the mass-shell relation $p_\m p^\m=0$. Hence, 
although in the
transverse plane there is a momentum transfer, as we observe from \eq{-4}, in 
the
forward direction this transfer is of ${\cal O}(\ve^2)$, and thus negligible to 
first order in
$\ve$. This is a specific feature of the lightcone coordinates, where the energy 
is proportional
to the square of the transverse momentum, instead of being linear in it. Observe 
also that,
since we have $\p_if\sim\f{1}{b}$, for large transverse
separations (compared to the Planck length) also the transverse momentum 
transfer is negligible.
We are then in the regime where the eikonal approximation is valid.
\section{An action principle for Planckian scattering}

In reference \rf{s} an action was proposed in terms of which an 
$S$-matrix can be constructed for the scattering process described in section 1. 
It was
obtained by a covariant generalisation of 't Hooft's action. The latter had been 
found by
considering the effect of the shockwave on the wavefunction of out-coming 
particles. For details
about this method we refer to \rf{gnato}. Here we show that, with a
further refinement we introduce in this section, one can find an equation of 
motion which
reproduces both the longitudinal and the transverse effects discussed 
previously.

The candidate action is\footnote{We have adopted the 
usual definition, in which a factor of $\@{-\det\,g_{\m\n}}$ or $\@{h}$ is 
included in the symbols $\e_{\m\n\a\b}$ and $\e_{ij}$, respectively, so that 
they transform as tensors. Although the $\e$-tensors appear here under the 
integral sign, we
actually have neglected a term with a derivative acting on them. We also leave 
out of
\eq{eom2} a term involving a certain Christoffel symbol. These terms, however, 
only show up when 
we depart from flat geometries.}
\be
S\smqu -\f{T}{2}\int\d^2\t\s\,\l(\@{h}h^{ij}g_{\m\n}\p_iX^\m\p_jX^\n 
+\f{1}{T}\,\e_{\m\n\a\b}X^\m P^\n\,\e^{ij}\p_iX^\a\p_jX^\b\r).
\le{action0}
Here, $X^\m$ is the field that describes the trajectories of particles, modified 
by the presence
of the momentum $P^\n$ of the other particles. \eq{action0} has the equation of 
motion
\be
\L X^\m\smqu\f{1}{2T\@{h}}\,\e^{\m\n\a\b}\,\e^{ij}\l(\p_iX_\a\p_jX_\b P_\n
+2\p_j(P_\n X_\b)\p_iX_\a\r),
\le{eom0}
which has a nonlocal solution. The first term under of \eq{eom0} can indeed be 
checked to 
reproduce the known effects, as we will do later. However, the second factor is 
rather
troublesome, for various reasons. One of them is that it gives extra 
contributions which do not
appear in the gravity calculation of section 1. This term comes from the 
variation of $X^\a$ and
$X^\b$ in the action. Indeed, \eq{action0}
cubic in the field $X^\m$, whereas the original 't Hooft action was only linear. 
We therefore replace \eq{action0} by the following action:
\be
S=-\f{T}{2}\int\d^2\t\s\,\l(\@{h}h^{ij}g_{\m\n}\p_iX^\m\p_jX^\n 
+\f{1}{T}\,\e_{\m\n\a\b}X^\m P^\n\,\e^{ij}\p_iX^\a_0\p_jX^\b_0\r).
\le{action1}
Here, $X^\m_0$ is the position of the test particle {\it before} the 
interaction, {\it i.e.},
unmodified by the shockwave. It is a fixed background, which in 't Hooft's 
calculation equals
$X^\m_0=(0,0,\s,\tau)$. Therefore, there is now just one $X^\m$ 
in the second term of \eq{action1} to be varied, coupled to an external source 
term:
\be
S=-T\int\d^2\t\s\@{h}\,\l(\f{1}{2}\,h^{ij}g_{\m\n}\p_iX^\m\p_jX^\n +J_\m 
X^\m\r),
\le{action2}
with
\be
J_\m=\f{1}{2T\@{h}}\,\e_{\m\n\a\b}\,W^{\a\b}_0\,P^\n,
\le{source}
and the orientation tensor $W^{\m\n}$ is defined in the following way: 
$W^{\m\n}_0\equiv\e^{ij}\p_iX^\m_0\p_jX^\n_0$. The source is thus a given 
function. The 
equation of motion is now
\be
\L X^\m(\t\s)=J^\m(\t\s),
\le{eom2}
and is solved by
\be
X^\m(\t\s)=X^\m_0(\t\s)+\int\d^2\t\s'\@{h}\,J^\m(\t\s')f\ts,
\le{soleom2}
where $f$ is the Green's function defined by
\be
\L f\ts=\f{1}{\@{h}}\,\dts.
\le{green1}
$X^\m_0$ is here a harmonic function of $\t\s$, depending on the boundary 
conditions on $X^\m$.

Now the action \eq{action2} reproduces the longitudinal shifts of the shockwave 
only. To include
the transverse effects as well, we have to do better. The validity of 
\eq{action2} is
limited
to the case of negligible transverse momenta, which corresponds to the limit 
$\ve\approx 0$. Now,
following \rf{s} and \rf{g}, our assumption is that a more general expression is 
provided by
replacing $X^\m_0$ by $X^\m$ in \eq{eom2}. We then get the equation of motion
\be
\L X^\m(\t\s)=\f{1}{2T\@{h}}\,\e^{\m\n\a\b}\,\e^{ij}\p_iX_\a\p_jX_\b P_\n.
\le{eom1}
As we will next show, this indeed gives the transvese effects.

Let us make one remark about \eq{action1}. One would be inclined to replace it 
by \eq{action0},
which does not depend on the background field $X^\m_0$. However, as argued at 
the beginning of
this section, when we calculate the equation of motion this action gives an 
extra term 
which does not yield the correct shift equations. Therefore, it cannot be 
assumed to be the
total action, but probably it has to receive new contributions that cancel the 
extra term in
\eq{eom0}. Because of the $\e$-tensor, we have not been able to find such a 
term. So we take
\eq{eom1} as our starting point for an improved theory, \eq{action1} 
corresponding to the
eikonal limit.

Now it is easy to check that, for the longitudinal variables $u$ and $v$, 
\eq{eom1} gives the correct result. We choose the gauge $X^\m=(u,v,x,y)$, and 
$\t\s\equiv\t x$. As in section 1, the in-going particles now only have momentum
$P_v$, and the out-coming particles 
momentum $P_u$. Then \eq{eom1} gives, for the $u$-component,
\be
u=u_0-\f{1}{T}\int\d^2\t x'\,P_v(\t x')\,\ftx.
\le{-11}
There is however a subtlety when we want to compare the solution of the 
equations of motion of the action \eq{-11} to the trajectories directly 
calculated from gravity. Our action was built up by $S$-matrix considerations of 
in-going and out-coming wavepackets. In particular, the second term of the 
action, \eq{action2}, which involves the momentum of the in-going particle, came 
up when considering out-coming trajectories {\it after} the shift, in other 
words, trajectories for which $v\geq 0$. All earlier trajectories remained 
unchanged. So, for the second term of \eq{-11} is only valid for trajectories 
$v\geq0$, we have to include a step function $\th(v)$ in order to compare it to 
the trajectories coming from the gravity calculation. Since we have the relation 
$\th(v)=\f{1}{2}+\f{1}{2}\,\mbox{sgn}\,(v)$, we can just replace the Heaviside 
function by $\f{1}{2}\,\mbox{sgn}\,(v)$, including the constant piece in $u_0$. 
This is allowed, since physically\footnote{The same can be said of the solution 
\eq{7} to the equations of motion, for which a Heaviside-function would have 
done as 
well.} what is of interest is only the relation \eq{3}, which is not modified by 
such redefinitions of $u_0$. Taking this into account (and doing the same for 
the in-going particles), we get for the $u$- and $v$-directions
\be
u=u_0-\f{1}{2T}\,{\mbox{sgn}}\,(v)\int\d^2\t x'\,P_v(\t x')\,\ftx,\nn
v=v_0+\f{1}{2T}\,{\mbox{sgn}}\,(u)\int\d^2\t x'\,P_u(\t x')\,\ftx,
\le{43}
which agrees with \eq{7}.

Next we check that also the transverse modes, which were not included in the 
original 't Hooft
action \rf{g9607}, automatically come out of the covariant equation of motion 
\eq{eom1}. We
concentrate on the motion in the $y-v$, $x-v$ planes. \eq{eom1} yields
\be
x^i=x^i_0+\f{1}{T}\int\d^2\t x'\,P_v(\t x')\,\p_iv(\t x')\ftx,
\le{50}
the index $i$ running over the values 2 and 3 in the transverse plane, 
$(x^2,x^3)=(x,y)$, and
so, upon taking the derive with respect to $v$ (notice that now $v$ is regarded 
as 
the affine parameter, just as in in section 1, so that the transverse variables 
become dependent on it) and changing variables, \eq{50} becomes
\be
\f{\p x^i}{\p v}=\f{1}{T}\int P_v\p_i v\,\p_v f=\f{1}{T}\int P_v\p_i\ftx.
\le{51a}
The dependence on the index $i$ thus corroborates the symmetry in the choice 
between $x$ and $y$. If $f$, for example, only depends on $y$, $x$ will be 
constant with respect to $v$, and viceversa.

To compare \eq{51a} to the trajectories \eq{7} we still have to include the 
sgn-function. However, when taking the derivative with respect to $v$, we would 
get a $\dt$-function contribution. Probably, the sgn-function, which must be 
included already on the right-hand side of $\L y=J^y$, goes under the integral 
sign of $y$ and therefore is not affected by the derivative; maybe at the end, 
taking 
$v$ as the affine parameter, independent of the transverse coordinates, it could 
be taken out of the integral sign, giving
\be
\f{\p x^i}{\p v}=\f{1}{2T}\,{\mbox{sgn}}\,(v)\int P_v\p_i\ftx,
\le{51b}
which is \eq{7}. This point, however, is still unclear and has to be studied in 
more detail. In the end we will make some remarks in this respect.

Equation \eq{51a} obviously also contains information about the transverse
momentum transfer that takes place during the collision. Writing
$p^\m=p\,\f{\p x^\m}{\p v}$ (as $v$ is the affine parameter
for the out-coming particles), \eq{51b} implies
\be
P^i(\t x)=\f{1}{T}\,P^v(\t x)\int\d^2\t x'\,P_v(\t x')\,\p_i\ftx.
\le{51c}
The momentum $P^v$ on the right-hand side of \eq{51c} is needed to cancel the 
factor of
$p$ that appears on the left-hand side of \eq{51c} when we replace $\f{\p 
x^i}{\p v}$ by $p^i$,
since for the trajectory of the out-coming particles {\it before} the collision 
holds
$x^\m=(0,v,\t x_0)$. This is the expression \eq{-5} found from kinematics of the 
process.

We thus see that the covariant equation of motion \eq{eom1} reproduces both the 
longitudinal and
the transverse effects of the shockwave, even though the original calculation 
only used the 
longitudinal effect. But as remarked, its solution  is nonlocal, which
means $X^\m$ does not transform as a scalar under general reparametrisations of 
$\t\s$. It,
however, does transform properly under the Lorentz group. In sections 5 and 6 we 
will see how
to obtain local expressions.

The second term of \eq{action2} being linear in $X^\m$ simplifies the $S$-matrix 
enormously,
since the path integral can be evaluated exactly. We are interested in an 
expression of the kind
\be
\bra {\mbox{out}}|{\mbox{in}}\ket={\cal N}\,\bra\e^{-iT\int J_\m X^\m}\ket,
\le{51d}
where $J^\m=J^\m_{\sm{in}}+J^\m_{\sm{out}}$ and the expectation value on the 
right-hand side
is taken with respect to the reference $S$-matrix element that is defined by
$\bra 1\ket={\cal N}^{-1}$ (see \rf{g9607}). We integrate over all fields $X^\m$ 
and possibly
(although this is not totally clear yet) over the metric
$h_{ij}$. For simplicity, we temporally gauge-fix the transverse metric to be 
flat, ignoring
ghosts and the anomaly in the path-integral measure. Evaluation of the path 
integral then gives
\be
\bra 
{\mbox{out}}|{\mbox{in}}\ket=\exp\l(-\f{iT}{2}\int\d^2\t\s\,\d^2\t\s'\,J_\m(\t\s
)J^\m(\t\s')
\fts\r).
\le{51e}
Notice that
\be
J_\m J^\m=-\f{1}{2T^2}\,P_\m P^\m W_{\a\b}W^{\a\b}-\f{1}{T^2}\,P_\m 
W^{\m\a}W_{\a\b}P^\b.
\le{51f}
In the limit that the transverse momentum of one of the particles can be 
neglected, for which
this action is valid, the second term of \eq{51f} vanishes, and we
get
\be
\bra 
{\mbox{out}}|{\mbox{in}}\ket=\exp\l(\f{i}{2T}\int\d^2\t\s\,\d^2\t\s'\,g_{\m\n}
P_{\sm{in}}^\m(\t\s)
P_{\sm{out}}^\n(\t\s')\fts\r).
\le{51g}
This is the nonlocal form of the 't Hooft $S$-matrix. Substituting the value of 
the momentum
distribution, \eq{4b}, the amplitude gives an expression which depends on the 
momentum of each
particle but also on the transverse location at which it enters or leaves the 
horizon. But one is
interested in an amplitude which only depends on the momenta, and hence should 
integrate over
all the possible points at which a particle can impinge the horizon. The 
amplitude \eq{51g} thus
becomes
\be
{\cal A}&=&\bra p_1^{\sm{out}}\cdots p_M^{\sm{out}}|p_1^{\sm{in}}\cdots 
p_N^{\sm{in}}\ket\nn
&=&\int\prod_{i,j=1}^{NM}\d^2\t x^i\,\d^2\t 
x^j\,\exp\l({\f{i}{2T}\sum_{i,j=1}^{NM}
\int\d^2\t x\,\d^2\t x'\,g_{\m\n}p_i^\m p_j^\n f(\t x^i-\t x^j)}\r)\nn
&=&\int\prod_{i,j=1}^{NM}\d^2\t x^i\,\d^2\t x^j\,|\t x^i-\t x^j|^{iG\,p_i\cdot 
p_j},
\le{51h}
using \eq{2}. Regularising this integral by dividing out the volume of the 
symmetry group \rf{gsw},
one gets the Koba-Nielsen generalisation of the Veneziano amplitude, as remarked 
in \rf{g87}.
The string constant is imaginary.
\section{Including the curvature of the horizon: relation to strings}

In this section we study the case that the horizon of the black hole where the 
particles interact is not flat, but has a nonvanishing (two-dimensional) Ricci 
scalar, and the effect of this on the $S$-matrix. In reference \rf{s} this 
possibility was not worked out. It was nevertheless argued that a dilaton term 
is expected to appear in the string theory action that effectively describes the 
dynamics at the horizon. But because one was unable to find it ---for it couples 
to the Ricci scalar on the world-sheet, which was zero in the Rindler 
approximation--- it was introduced by hand. We will see that this 
term naturally appears if one considers curved metrics from the beginning.

In order to do this we return to the equations for the shift, equation \eq{9}, 
that the function $f$ must satisfy at the plane $v=0$. Although with a 
particular choice of
coordinates, these conditions include the effect of the transverse curvature. We 
now look for a
more
general form of this expression, which can be incorporated into the $S$-matrix. 
This is readily
found if one considers that the second term on the left-hand side of \eq{9} is a 
relic of the
two-dimensional Ricci-tensor. 
Indeed, according to the calculation outlined in Appendix B, the Ricci tensor of 
the vacuum metric \eq{8} equals
\be
R_{ij}=R_{ij}^{(2)}-h_{ij}\f{g_{,\,uv}}{A},
\le{51}
where $R_{ij}$ is the transverse part of Ricci tensor obtained from the full 
four-dimensional Riemann tensor, see \eq{B3}, and $R_{ij}^{(2)}$ is the 
Ricci tensor corresponding to the two-dimensional metric $h_{ij}$. Now we notice 
that $R_{ij}$ satisfies the vacuum Einstein equations\footnote{It corresponds to 
the metric before we apply the shift, see the Appendix B.}, and hence we have 
that solutions of this equation must satisfy
\be
R^{(2)}=\f{2}{A}\,g_{,\,uv}.
\le{53}
Therefore we can write the second of equations \eq{9} as
\be
\l(\L-\f{1}{2}\,R^{(2)}\r)f&=&\f{1}{\@{h}}\,\dt^{(2)}(\t x-\t x_{\sm{0}}),
\le{green}
$\L$ and $R$ being calculated in the same metric $h_{ij}$.

It is easy to include this extra contribution in our action. To this purpose we 
apply the
philosophy of \rf{s}. When the curvature term was zero, it was argued that the 
action
reproducing \eq{green} is given by \eq{action1}. The
first term gives the free propagator, which is the first summand on 
the right-hand side of \eq{green}. The second term of the action is an 
interaction term, which provides the source on the right-hand side of 
\eq{green}. Now it is 
clear that the term involving the curvature should be a mass term, and therefore 
the action becomes
\be
S=-\f{T}{2}\int\d^2\t\s\,\l(\@{h}h^{ij}g_{\m\n}\p_iX^\m\p_jX^\n
+\f{1}{2}\,\@{h}R^{(2)}X_\m X^\m 
+\f{1}{T}\,\e_{\m\n\a\b}X^\m P^\n\,\e^{ij}\p_iX^\a_0\p_jX^\b_0\r).
\le{action}
The equation of motion is
\be
(\L-\f{1}{2}\,R^{(2)})\,X^\m= \f{1}{2T\@{h}}\,\e^{\m\n\a\b}\,W_{\a\b}^0\,P_{\n},
\le{eom}
which generalises \eq{green}. This is of course \eq{eom1} with the extra 
curvature 
term, and reminds of a focusing theorem. The general solution to this equation 
is 
again \eq{soleom2},
\be
X^\m(\t\s)=X^\m_0(\t\s)+\f{1}{2T}\int\d^2\t\s'\,\e^{\m\n\a\b}\,W_{\a\b}^0P_\n\, 
f\ts,
\le{soleom}
but with $f$ satisfying \eq{green} instead of \eq{green1}.

We now go to the string interpretation of \eq{action}. We temporarily bypass the 
problem of
the extra contribution to the equation of motion in \eq{eom0}, assuming that if 
one
is able to construct an action that allows for transverse momenta and gives 
\eq{eom1} back, the
$X^\m_0$ in \eq{action} will be replaced by $X^\m$, with additional terms that 
cancel the extra
contribution to \eq{eom0}.

The first term in \eq{action} is the
Polyakov term which describes the free propagating string, and, as remarked in 
\rf{s}, the
third looks formally like an antisymmetric tensor field $B_{\m\n}$ coupled to a 
Wess-Zumino term.
But in this model the antisymmetric tensor field is determined by the position 
and momenta of
the particles, and reads
\be
B_{\m\n}(X)=\e_{\m\n\a\b}\,X^\a P^\b ,  
\le{41}
from which a field-strength is derived which is just the dual of the momentum,
\be
H_{\m\n\a}=3\,\e_{\m\n\a\b}\,P^\b.
\le{42}

The second term of the left hand side of equation \eq{eom} had, up to a 
factor of $\f{1}{2}$, already been found in reference \rf{erik}. 
However, Verlinde and Verlinde found the first and second terms of the action 
\eq{action} from rather different considerations, namely by a separation of the 
Einstein action in a strongly coupled and a weakly coupled piece, corresponding 
to the two physical length scales involved in the problem (a longitudinal and a 
transverse one)(see section 5 for a connection with this idea). But here it 
arises altogether
with a term that can possibly be interpreted as an antisymmetric tensor 
background. Due to
this formal similarity with the world-sheet action of string theory, it is 
tempting to regard the
curvature term as a dilaton contribution:
\be
S_{\sm{dil}}=-T\int\d^2\t\s\,\@{h}R^{(2)}\Phi(X),
\le{43}
where the dilaton is given by
\be
\Phi(X)=\f{1}{2}\,g_{\m\n}X^\m X^\n.
\le{44}
If this is correct, then the content of the effective theory induced at the 
horizon is exactly
the same as that of string theory, including the antisymmetric tensor and the 
dilaton, with
explicit expressions for these fields. 
However, as one can see from \eq{43}, the dilaton comes in with the same power 
in the string coupling as the Polyakov term, so that its contribution in the 
path integral is of the same order as that of the free term. Therefore, in 
general, conformal
invariance will be destroyed, and it will arise only for the massless case, 
which corresponds
to flat shockwaves.
\section{A generalised Heisenberg principle}
One would like to know what the Hilbert space structure of this model is, that 
is, how operators
act on Hilbert space and what their eigenvalues are. The quantisation of this 
model for the
cases where the transverse momenta 
are exactly zero and the fluctuations of the shockwave are small was already 
discussed in \rf{s}. Now we try to deal with the more general case.

We first consider the action of the operators $\^P^\m$ and $\^X^\m$ on the state 
vectors $|P\ket$
and $|X\ket$. We obviously have
\be
\^P^\m|P\ket=P^\m|P\ket;\nn
\^X^\m|X\ket=X^\m|X\ket.
\le{79}
Furthermore, the states $|P\ket$ and $|X\ket$ as introduced in the path integral 
were related by
a Fourier transformation \rf{s}, so that we have
\be
&[\^X^\m(\t\s),\^P^\n(\t\s')]=ig^{\m\n}\,\dts.
\le{80}
Indeed, when constructing the $S$-matrix, one used the operator $\^P^\m$ to 
generate the shifts
of the wavefunction that reproduce the shockwave effect when two particles cross 
each other.
\eq{80} therefore agrees with the interpretation of $X^\m$ as the eigenvalue of 
a position
operator, and of $P^\m$ as that of the momentum operator (more properly, the 
distribution of
momentum along the shockwave).

Furthermore, we have to require that the results of measurements of the 
positions and momenta of
the particles obey the equation of motion \eq{eom}. This holds both for the 
in-going and for the
out-coming particles. Therefore, we have to require that the eigenvalues of 
these
operators obey this equation as well. The only way we see to do this is by 
identifying the
corresponding operators in Hilbert space (notice that this connects the Hilbert 
spaces of
{\it different} particles, since \eq{eom} relates the trajectory of one particle 
to the momentum
of the other). So, following \rf{g9607}, we propose \eq{eom} to be also a 
relation between the
{\it operators} $X^\m$ and $P^\m$ acting on the Hilbert spaces of the different
particles\footnote{By an abuse of notation, we use the same symbol for the 
operators $\^X^\m$,
$\^P^\n$ as for the corresponding eigenvalues $X^\m$ and $P^\n$.}. This imposes 
a strong
constraint on the space of states, for it identifies position states with 
momentum states in a
certain way. Later on we will see that also the in- and the out-Hilbert spaces 
are the same.

Equation \eq{80}, combined with the operator analog of \eq{eom}, now leads (as 
shown in \rf{s})
to the following commutation relation:
\be
&[X^\m(\t\s),X^\n(\t\s')]=-\f{i}{2T}\,\e^{\m\n\a\b}\,W_{\a\b}\fts.
\le{81}
One has to keep in mind that the two operators $X^\m$ here correspond to 
different (bundles of)
particles.

However, it was already remarked that in general, and in particular when we 
consider departures from a flat shock-wave, since the right-hand side of \eq{81} 
becomes itself an operator, equation \eq{80} will receive ${\cal O}(\ve)$ 
corrections, which, nevertheless, do not modify \eq{81} to first order in $\ve$.
We now investigate these corrections. We anticipate that the fact that we are 
dealing with fields that depend on the world-sheet variables $\t\s$ in a 
nontrivial fashion makes things very complicated, and we were not able to find 
an analog of \eq{81} that holds for arbitrary shock-waves. Yet, considering the 
almost-flat case still gives us a lot of information.

Our Ansatz (see appendix C) for the extra term that appears in \eq{80} is the 
following
\be
&[X^\m(\t\s),P^\n(\t\s')]=i\l(g^{\m\n}+A^{\m\n}\r)\dts,
\le{Amunu}
where
\be
A^{\m\n}=-\f{1}{T\@{h}}\,\e^{\m\n\a\b}
\e^{ij}\p_iX_\a\int\d^2\t\s''\,P_\b(\t\s'')\,\p_jf(\t\s-\t\s'').
\le{AAmunu}
The general form of this expression can be calculated from the Jacobi identity, 
although ---due
to the nonlocality--- one does not get the correct $\t\s$-dependence from this 
(in
appendix C we show where the problem comes from). However, one can justify the 
guess \eq{Amunu}.
At the end of this section we will see that, combined with the equations of 
motion,
\eq{Amunu} represents the momentum transfer and makes the momentum operator be 
the generator of
translations, as one would expect in a field theory, and in
section 6 we will perform an independent check of this expression by 
dimensionally reducing it
and comparing it with the 2+1 dimensional case, where the nonlocality problems 
can be avoided.

Notice that for large impact parameter, $\f{1}{b}\ll 1$ in Planck units, the 
extra term of
\eq{Amunu} can be neglected.

The modification of the canonical commutation relation \eq{80} in the presence 
of gravitational interactions has also been predicted (although in different 
contexts) by several authors \rf{uncertainty}\rf{max}:
\be
\D x \D p\geq\f{\hbar}{2}+{\cal O}\l(\f{\Pl^2\,p_{\sm{in}}}{b}\r).
\le{92}
To understand the physical meaning of this, we first draw an analogy with string 
theory. The  action
of a string interacting with an
antisymmetric tensor field background  $B_{\m\n}$ can, in complex coordinates, 
be written as
\be
S=-\f{T}{2}\int\d z\d\ov z\l(g_{\m\n}+B_{\m\n}\r)\ov\p X^\m\p X^\n.
\le{93}
There are two conserved quantities, let us call them $\Pi_\m$ and $\ov{\Pi}_\m$. 
 Choosing
$\ov z$ to be the time variable in these complex ``light-cone" coordinates, the 
canonical
momentum is
\be
\Pi_\m=\f{\dt S}{\dt\ov\p X^\m}=-(g_{\m\n}+B_{\m\n})\p X^\n 
\equiv(g_{\m\n}+B_{\m\n})P^\n.
\le{94}
It obeys the equal-time commutation relation
\be [X^\m(z,\ov z),\Pi^\n(z',\ov z)]=ig^{\m\n}\,\dt(z-z').
\le{94b}
In analogy with our model (where $\ve$ was the expansion parameter), let us 
assume
that the field $B_{\m\n}$ is proportional to some very small parameter  $\eta$. 
Then an
Ansatz consistent with \eq{94} to first order in $\eta$ is
\be
P_\m=(g_{\m\n}-B_{\m\n})\Pi^\n+{\cal O}(\eta^2).
\le{95}
This variable is the ``flat-space" canonical momentum, that is, the generator of 
translations in the absence of the $B$-field. It satisfies
\be
[X^\m(z,\ov z),P^\n(z',\ov z)]=i(g^{\m\n}+B^{\m\n})\,\dt(z-z').
\le{96}
So the presence of the $B$-field modifies the canonical
momentum and  also the commutation relation between the would-be
canonical momentum $P^\m$, and  $X^\m$.

A similar thing is happening in our model. The commutator \eq{Amunu} is not 
the canonical one, which suggests that we have to look for another variable 
which 
is the canonical momentum. This brings us to the following definition:
\be
P^\m_{\sm{can}}(\t\s)\equiv (g^{\m\n}+A^{\m\n})\,P_\n(\t\s).
\le{Pcan}
One can easily check that taking the commutator of this operator with $X^\m$ 
gives
\be
[X^\m(\t\s),P^\n_{\sm{can}}(\t\s')]=ig^{\m\n}\dts+{\cal O}(\ve^2).
\le{97}
Notice that, although $A^{\m\n}$ is itself also an operator, to first order in 
$\ve$ this fact does not affect the canonical commutation relation \eq{97}, 
since, for the in-going particles, $A^{\m\n}$ is built up of out-coming 
operators, 
which, to zeroth order in $\ve$ (since $A^{\m\n}$ is itself already 
${\cal O}(\ve)$), commute with the position operator of the in-going particles.

The generalised commutator \eq{Amunu} has a simple interpretation if we go 
back to the shock-wave picture underlying the action \eq{action}. Before the 
interaction takes place, the different coordinates are not yet coupled 
through the equations of motion \eq{eom} or \eq{7}, {\it i.e.}, the out-coming 
fields $X^\m$ do not yet depend on $\t\s$. However, {\it after} the 
interaction, the coordinates have got mixed and $P_u$ can, for example, generate 
sideways
displacements as well. So it is natural to identify the 
canonical momentum $P^\m_{\sm{can}}$ with the momentum before the 
interaction\footnote{However,
the existence of such an operator again violates the Jacobi identity. Probably 
there is here
some subtlety with the time variable involved, since $P_{\sm{can}}^\m$ is 
defined for times
before the collision, when the coordinates $X^\m$ still commute among 
themselves.}, 
which we denote by $P^\m_0$, and $P^\m$ with the momentum after the shockwave. 
The latter
describes the momentum transfer, and can be
seen to be a measure for the recoil of the particles. This aspect had not been 
taken into account
in earlier works. This holds both for the in-going and the out-coming particles. 
Since to
zeroth order in $\ve$ both momenta agree, we will drop the subscript $0$ in 
expressions which are
already ${\cal O}(\ve)$. Although $P^\m$ is a non-canonical operator, when 
writing \eq{Amunu} out
in components we will conclude that it generates translations in the sense of 
field theory.

This definition turns out to be very powerful since it allows us to calculate 
expressions for
the momentum transfer after the shock-wave interaction. An attempt to derive an 
algebra that
would hold even if there is this recoil was pioneered in \rf{g94}.

So we now go to Minkowski space to study the momentum transfer between the 
in-going and the outcoming particles. We allow the momentum distributions to be 
in any direction, $P^{0,\,{\sm{in}}}_\m=(P_u,P_v,P_x,P_y)_{\sm{in}}$ and 
$P^{0,{\sm{out}}}_\m=(P_u,P_v,P_x,P_y)_{\sm{out}}$. The only constraint is that 
the particles have to be massless. So, in particular, we do not require the 
transverse momenta to be zero, differently from the standard approach.
For convenience, we split the indices in a longitudinal and a transverse part, 
$\m=(\a,i)$, denoted by (the first few) Greek and Latin letters, respectively. 
For the transverse
parts, we again have the gauge $x_0^i=\s^i$.
With the above discussed definition of momentum before and after crossing the 
shockwave, we can invert \eq{Pcan} to
\be
P^\m=(g^{\m\n}-A^{\m\n})P_\n^0+{\cal O}(\ve^2),
\le{98}
both for the in- and for the out-coming particles.

We now can calculate the  transverse momentum transfer that takes place
during the  interaction\footnote{Remember that we dropped the $0$ on the
right-hand side.}:
\be
P_i(\t\s)&=&\f{1}{T}\, P_v(\t x)\int\d^2\t\s'
\,P_u(\t\s')\,\p_i\fts -\f{1}{T}\, P_u(\t\s)\int\d^2\t\s'\,
P_v(\t\s')\,\p_i\fts\nn
&\equiv&\f{1}{T}\,\e^{\a\b}P_\a\int\d^2\t\s'\,P_\b\,\p_i\fts.
\le{99}
In the last line we have used covariant notation. The labels ``in" or ``out" are 
left out,
since it should be clear that the  operator on the
right-hand side of \eq{100} which is evaluated at the same point  $\t\s$ as 
$P_i(\t\s)$,
corresponds to the same particle, whereas the  operators which are integrated 
over give
the contributions of the other  particles. Notice that even if the
initial transverse momentum is zero, so that the two  particles have a
head-on collision, after the interaction it will not vanish  anymore.
The two particles will `spin' around each other for a very short time. 
This is in accordance with the picture of section 2. Equation \eq{99} is (up to
a minus sign) indeed in complete agreement with \eq{-4} and \eq{51c},
which were calculated from kinematical  considerations and by
considering the transverse trajectories, respectively. If the ``hard
particle" has only longitudinal initial momentum, so that the  equation
of motion \eq{-11} (written in the 2-2 splitting of space-time) for the 
longitudinal part can
be used,
\be
X^\a=\f{1}{T}\int\e^{\a\b}P_\b\,f,
\le{100b}
we get
\be
P_i(\t\s)=P_\a\p_iX^\a.
\le{100}
Then, \eq{99} can also be written as
\be
P_i(\t\s)&=&\f{\e}{T}\,P^\a(\t\s)\int\d^2\t\s'\,P_\a(\t\s')\,\p_if(\t\s-\t\s'),
\le{100c0}
where $\e=1$ if $P_i$ is an operator corresponding to the in-going particles and 
$\e=-1$ for the
out-operators. This is 't Hooft's sign convention for momenta \rf{g9607}, where
all in-going momenta are defined to be positive, and out-coming momenta
to be negative. Indeed, if initially the in-going particles only have
momentum $P_v$, and the out-coming ones only momentum $P_u$, \eq{100c0}
gives us
\be
P_i^{\sm{in}}(\t\s)&=&\,\,\;\f{1}{T}\,P_v^{\sm{in}}(\t\s)\int\d^2\t\s'\,P^{\sm{o
ut}}_u(\t\s')
\,\p_i\fts\nn
P_i^{\sm{out}}(\t\s)&=&-\f{1}{T}\,P_u^{\sm{out}}(\t\s  )\int\d^2\t\s'\,
P_v^{\sm{in}}(\t\s') \,\p_i\fts,
\le{100c}

Let us now consider the longitudinal momentum transfer. Using \eq{98}, this 
amounts to
\be
P_u(\t\s)&=&-\f{1}{T}\,P^i\int\d^2\t\s'\,P_u\,\p_if 
+\f{1}{T}\,P_u\int\d^2\t\s'\,
P^i\,\p_if,\nn
P_v(\t\s)
&=&+\f{1}{T}\,P^i\int\d^2\t\s'\,P_v\,\p_if 
-\f{1}{T}\,P_v\int\d^2\t\s'\,P^i\,\p_if.
\le{101}
In covariant notation,
\be
P^\a=\f{1}{T}\,\e^{\a\b}\l(P_\b\int P^i\,\p_if-P^i\int
P_\b\,\p_if\r).
\le{102}
In the case that the ``hard particle" moves along one of the lightcone 
directions without
transverse momentum, the first term vanishes, and
\eq{102} can be written as
\be
P^\a(\t\s)=-P^i(\t\s)\,\p_iX^\a(\t\s).
\le{103}
Notice that this expression does vanish if the initial transverse momentum of  
the test
particle is zero. So, although for initial zero transverse momentum there is 
still a
transverse momentum transfer, in the longitudinal plane this transfer is zero to 
first order
in $\ve$. This agrees with the Verlinde  idea \rf{erik} of separating the
gravitational action into two sectors, one  corresponding to the longitudinal 
and the other
corresponding to the transverse modes, with two different coupling constants. 
They argued
that, for low momentum transfer ({\it i.e.}, for large transverse separations), 
the transverse
modes become  classical, and only the longitudinal ones have to be treated 
quantum  mechanically.
This corresponds to 't Hooft's analysis. From  \eq{103} we learn indeed that the 
longitudinal
momentum transfer is negligible to first order in $\ve$ if the transverse 
momentum
is negligible, so that the eikonal approximation can be used. However, we also 
find that even if the
initial transverse momenta are zero, when the impact parameter becomes small, 
there is a
transverse momentum transfer, and the transverse modes have to be treated
quantum mechanically as well. The eikonal approximation can then not be trusted 
anymore.
Put differently, the transverse modes can be treated classically if the momentum
transfer is small, but that (transverse) transfer is not small for Planckian 
longitudinal
momenta, and as soon as the distance between the particles is comparable to the 
Planck length.
The formalism developed here, although not satisfactory for the reasons 
explained, seems to
hold even when the transverse momenta are not negligible, and beyond the eikonal 
approximation.

Let us now take a closer look at the commutation relation \eq{Amunu},
\be
[X^\m(\t\s),P^\n(\t\s')]=iG^{\m\n}\,\dts,
\le{104}
where the ``generalised metric" is defined as
\be
G^{\m\n}\equiv g^{\m\n}+A^{\m\n}.
\le{105}
Writing \eq{104} out in components, the relevant equations are
\be
{}[u(\t\s),p_i(\t\s')]&=&i\p_iu\,\dts;\nn
{}[v(\t\s),p_i(\t\s')]&=&i\p_iv\,\dts;\nn
{}[x^i(\t\s),p_u(\t\s')]&=&i\p_ux^i\,\dts;\nn
{}[x^i(\t\s),p_v(\t\s')]&=&i\p_vx^i\,\dts,
\le{106}
where we remind that the equations of motion give $x^i=x^i(u(\t\s),v(\t\s))$. In
covariant notation, the commutators \eq{106} become
\be
{}[x^\m(\t\s),p_\n(\t\s')]&=&i\p_\n x^\m(\t\s)\dts.
\le{107}
As expected, the operator $p^\m$ generates translations in the direction $x^\m$. 
In quantum
mechanics, the different coordinates are independent of each other, and 
therefore the
right-hand side of \eq{107} reduces to $i\dt^\m_\n\,\dts$, the canonical 
commutator. But in
this case, since we have a two-dimensional field theory, the longitudinal and 
the transverse
coordinates become mutually dependent, so that \eq{107} is nonvanishing even for 
different
indices $\m$ and $\n$ (notice that, for $\m\not=\n$, it is nonzero only if one 
index is a
world-sheet index $i$, and the other a target-space index $\a$). So $p^\m$ 
generates
translations as in (a two-dimensional) field theory.

One can also get an algebra for the commutator of the $p^\m$'s among themselves. 
This cannot be
done covariantly from the equation of motion \eq{eom} because of the 
$\e$-tensor. However, one
can do it in the coordinate system we are using. With some algebra, one comes to 
(the operators
referring all to the in- or all to the out-states)
\be
{}[p_\a(\t\s),p_i(\t\s')]&=&ip_\a(\t\s')\,\p_i\dts;\nn
{}[p_i(\t\s),p_j(\t\s')]&=&ip_i(\t\s')\p_j\dtx+ip_j(\t\s)\p_i\dts,
\le{108}

Now we can also obtain an algebra that relates the in- and the out-operators. 
Using \eq{108}, we
get:
\be
{}[p_v^{\sm{in}}(\t\s),p_i^{\sm{out}}(\t\s')]&=&-iT\,\p_iu(\t\s')\,f^{-1}\ts;\nn
{}[p_u^{\sm{out}}(\t\s),p_i^{\sm{in}}(\t\s')]&=&iT\,\p_iv(\t\s')\,f^{-1}\ts;\nn
{}[p_i^{\sm{in}}(\t\s),p_j^{\sm{out}}(\t\s')]&=&-iT\,\p_iv(\t\s)\,\p_ju(\t\s')\,
f^{-1}\ts\nn
&+&\f{i}{T}\,p_v^{\sm{in}}(\t\s)\,p_u^{\sm{out}}(\t\s')\,\p_jf\ts.
\le{110}
In reference \rf{g9607} it was not possible to find correct expressions for 
\eq{110} because, in
particular, the expected expression for the last of \eq{110} did not satisfy the 
Jacobi identity
when combined with \eq{108}. One can check that the above expression satisfies 
it exactly.

The algebra \eq{110} is very nonlocal and, furthermore, nonlinear, and so not 
very useful.
It, however, can be significantly simplified by defining the total momentum
\be
P_\m=\int\d^2\t\s\,p_\m(\t\s).
\le{111}
Now one can check that for these variables \eq{110} leads to correct 
expressions:
\be
{}[p_i^{\sm{in}}(\t\s),P_j^{\sm{out}}]&=&i\p_jp_i^{\sm{in}}(\t\s);\nn
{}[p_i^{\sm{out}}(\t\s),P_j^{\sm{in}}]&=&i\p_jp_i^{\sm{out}}(\t\s);\nn
{}[p_\a^{\sm{in}}(\t\s),P_i^{\sm{out}}]&=&i\p_ip_\a^{\sm{in}}(\t\s);\nn
{}[p_\a^{\sm{out}}(\t\s),P_i^{\sm{in}}]&=&i\p_ip_\a^{\sm{out}}(\t\s),
\le{112}
the total transverse momentum generating translations, which is the usual 
situation in field
theory. However, a local expression is lacking.

One can check that the algebra between the transverse in-operators themselves or 
the
out-operators themselves is similar to \eq{112}. However, we do not expect the 
theory to have
two different generators of transverse translations. So we must have
\be
P_{\sm{in}}^i=P^i_{\sm{out}}.
\le{113}
Integrating \eq{99} we indeed see that this is the case. The same holds for the 
lightcone
directions, as one sees from equation \eq{101}). Therefore, for the integrated 
momentum
operators we get the constraint
\be
P^\m_{\sm{in}}=P^\m_{\sm{out}}.
\le{114}
Recalling that these operators correspond to the momentum transfer, this clearly 
implies that
there is momentum conservation during the collision. Since this holds for the 
operators acting
on any state, as a condition on Hilbert space this constraint is very strong: it 
means that the
Hilbert space of the in-going particles is the same as that of the out-coming 
particles! This
suggests that it is enough to look at Hawking radiation to
know what the structure of the inner of the black hole is.

Some of these results had already been predicted in \rf{g94}, but here they come 
naturally out
of the formalism. Furthermore, we get \eq{110} and \eq{112} for free. Of course, 
it remains
to be clarified why in \eq{Amunu2} we get an extra term. As said before, the 
reason must be the
nonlocal dependence of $X^\m(\t\s)$ on $\t\s$. Therefore, we do not
pretend to have proven \eq{AAmunu}. But the fact that it means that the momenta 
generate
translations in the different directions and gives the correct momentum transfer 
(also obtained
in section 2 from kinematics), suggests that this term will stay there when one 
has
a theory with better defined variables and a local algebra, instead of having to
``guess" the result, as we did in deriving its presence from consistency of the 
Jacobi
identity. Therefore, we have taken it as our starting point. Actually, in the 
next section we
perform another check that this expression gives the right result. So, although 
the picture is
not complete and the formalism is non-local, there is strong evidence that in 
the presence of
gravitational interactions, Heisenberg's principle has to be modified similarly 
to \eq{104}.
Notice furthermore that \eq{104} {\it does} give a local algebra, \eq{107}, 
after using the
equations of motion.
\section{Reduction to 2+1 dimensions}

The $S$-matrix Ansatz becomes much simpler if we consider a three-dimensional 
world, or, better
said, a world with one of the space-time (and world-sheet) directions 
compactified on a circle
of radius $R_3$, say $0\leq \s_2=y\leq R_3$. To derive the $S$-matrix in this 
$2+1$ dimensional
world\footnote{For another approach, see \rf{zeni}.}, we depart from the 
$3+1$-dimensional
action \eq{action} and then dimensionally reduce it, like in Kaluza-Klein 
theories. The fields
and momenta are assumed to be independent of this fourth coordinate. 
Correspondingly, at the
path-integral level the integration over $y$ will drop out. The action 
\eq{action} then reduces
to
\be
S=-\f{T_3}{2}\,\int\d\s\l(g_{\m\n}\,\p x^\m\p x^\n 
+\f{2}{T_4}\,\e_{\m\n\a}\,x^\m p^\n\p x_0^\a\r).
\le{2.9}
We have defined $\s\equiv\s_1$, $\p\equiv\f{\p}{\p\s}$, and 
$\e_{\m\n\a}\equiv\e_{\m\n\a\s_2}$. 
Notice that we have a renormalized overall coupling, $T_3\equiv T_4R_3$, which 
gives us an effective $2+1$-dimensional gravitational constant:
\be
G_3=\f{G_4}{R_3}.
\le{2.10}
However, the relative coupling in the second term, $\f{1}{T_4}$, does not 
change. 
This is the coupling that will appear in the equations of motion, and these 
therefore involve
the radius of the periodic dimension\footnote{From now on we will take $G_3$ as 
the ``true"
gravitational coupling in the lower dimension, and will therefore denote $T_3$ 
by $T$.}. From
\eq{2.10} we see that, for fixed $G_4$, if we send $R_3$ to zero the 
three-dimensional
gravitational constant diverges, whereas for $R_3\rightarrow\infty$ (the 
compactified dimension
being ``lifted"), $G_3$ becomes very small.

If the action \eq{action} appeared as a string theory action, \eq{2.9} looks 
like 
the action of a relativistic particle, whose motion would be parametrized by a 
(spacelike) variable $\s$. The equation of motion of \eq{2.9}, with $x^\m_0$ 
replaced by $x^\m$,
is
\be
\p^2x^\m=\f{R}{T}\,\e^{\m\n\a}\,p_\n\,\p x_\a.
\le{2.11}
This, again, is our starting point to include transverse effects.
Now, as in the preceding, we postulate the canonical commutation relation 
between $x^\m$ and
$p^\m$,
\be
[x^\m(\s),p^\n(\s')]=ig^{\m\n}\dt(\s-\s').
\le{2.13}
Furthermore, when quantizing this system, we have to take into account the 
equation of motion \eq{2.11}, which gives the leading contribution to the path 
integral. So, as in the $3+1$-dimensional case,  commuting \eq{2.11} with $x^\n$ 
yields
\be
[x^\m(\s),x^\n(\s')]=-\f{iR}{T}\,\e^{\m\n\a}\,\p x_\a(\s')\,f(\s-\s'),
\le{2.14}
which implies
\be
[\p x^\m(\s),\p x^\n(\s')]=\f{iR}{T}\e^{\m\n\a}\,\p x_\a\,\dt(\s-\s').
\le{2.15}
Notice that this is an SO(2,1) invariant algebra. It is local, and therefore 
seems to be a
good starting point for a covariant theory in $2+1$ dimensions. The same algebra 
has been
obtained by 't Hooft using another method \rf{g}.

It is useful to define variables that anyhow have a better dependence on the 
location at the
horizon. 't Hooft has defined the following variables:
\be
a^\m_A\equiv\int_A\d^2\s\,\p x^\m=x^\m(A_1)-x^\m(A_0).
\le{2.15b}
These have the nice property:
\be
[a^\m_A,a^\n_A]=\f{iR}{T}\,\e^{\m\n\a}\,a_\a^A,
\le{2.15c}
where for simplicity we restrict ourselves to the same region $A$ in $\s$-space. 
't Hooft has
remarked that this gives a time variable that is quantized in units of 
$t_{\sm{Pl}}/R$.
Another useful quantity is the total momentum flowing through $A$,
\be
p^\m_A\equiv\int_A\d\s\,p^\m(\s).
\le{2.16}
The canonical commutator then becomes
\be
[a^\m_A,p^\n_A]=ig^{\m\n}.
\le{2.16b}
But, as in the four-dimensional case, equation \eq{2.16b} together with 
\eq{2.15c} does not
satisfy the Jacobi identity, since the operators $x\m$ and $p^\m$ are no longer 
independent of
each other. One can easily check that, in order to satisfy the Jacobi identity, 
one has to
add an extra term to \eq{2.16b}. Analogously to the $3+1$-dimensional case, 
\eq{2.16b} is
modified in the following way:
\be
[a^\m_A,p^\n_A]=iG^{\m\n},
\le{2.17}
with the ``generalized metric"
\be
G^{\m\n}=g^{\m\n}-\f{R}{T}\,\e^{\m\n\a}\,p_\a^A.
\le{2.18}

Because of the equation of motion \eq{2.11}, the algebra \eq{2.17} can be
expressed in terms of the $a^\m$'s alone. We first note the following relations
(we use the notation $u_A\equiv a^u_A$, $v_A\equiv a^v_A$, etc.):
\be
p_v^A&=&\f{T}{R}\l(\f{\p x}{\p v}\r)_A=-\f{T}{R}\,(\p u)_A\nn
p_u^A&=&-\f{T}{R}\l(\f{\p x}{\p u}\r)_A=\f{T}{R}\,(\p v)_A.
\le{2.19}
Now we can write \eq{2.17} out in components:
\be
{}[u_A,p_x^A]&=&i(\p u)_A\nn
{}[v_A,p_x^A]&=&i(\p v)_A\nn
{}[x_A,p_u^A]&=&i\l(\f{\p x}{\p u}\r)_A\nn
{}[x_A,p_v^A]&=&i\l(\f{\p x}{\p v}\r)_A,
\le{2.20}
while all other commutators remain unchanged. Hence, $p^\m$ is the generator of 
translations in
the $a^\m$-direction, as expected.

The algebras \eq{2.15c} and \eq{2.17} do not have the $\t\s$-dependence anymore. 
Therefore
they do not suffer the problems mentioned before. One can therefore think of 
using them as a test
for our proposal \eq{AAmunu}. Upon dimensional reduction of \eq{Amunu}, one gets
\be
{}[x^\m(\s),p^\n(\s')]=i\l(g^{\m\n}-\f{R}{T}\,\e^{\m\n\a}\int\d\s''\,p_\a(\s'')\
p f(\s-\s'')\r)
\dt(\s-\s').
\le{2.21}
This equation still has a bad $\s$-dependence. But now if we go over to the 
variables $a^\m_A$
and $p^\n_A$, using the fact that in one dimension 
$f(\s-\s')=\f{1}{2}\,|\s-\s'|$, one can
check that \eq{2.21} exactly gives \eq{2.17}. In the same way, one can check 
that direct
dimensional reduction of the four-dimensional commutator \eq{81} gives 
\eq{2.15}, with no
need to integrate. One could think that the agreement is due to the fact that 
the
$2+1$-dimensional action
has been obtained by dimensional reduction of the $3+1$-dimensional one. This, 
however, is not
exactly the case. The $2+1$-dimensional algebra \eq{2.15c} does not have the 
nonlocality
problems of its higher dimensional analogous, and is closed. \eq{2.17} was 
derived directly
from this
local algebra. What is nontrivial is that when reducing the algebras \eq{81} and 
\eq{Amunu},
the problems of the four-dimensional case disappear. It could well have happened 
that we
got different expressions from \eq{2.15c} or \eq{2.18}, as is probably the case 
with
\eq{Amunu2} (although we did not check this). This
further advocates for \eq{Amunu} as being correct in our gauge.
\section{Discussion and conclusions}
Although at a very rudimentary stage, some evidence has been found in favour of 
the holographic
hypothesis, according to which the relevant degrees of freedom of a black hole 
sit at the
horizon. In particular, we find that the in- and out-Hilbert spaces are the 
same. Via the
shockwave model for the black
hole, we indeed get a field theory living on the horizon. This field theory has 
(or is defined
by) in- and out-operators
which add momentum or generate translations, and satisfy an algebra that can be 
brought to a
local form. The action describing the dynamics of the surface has an
interpretation as a string theory action, with background fields that are given 
by the momenta
and positions of
the particles. In particular, we have found an antisymmetric tensor background 
and a dilaton
term, which breaks conformal invariance at the classical level. The scattering 
amplitude one
obtains is the well-known Veneziano formula already found in \rf{g87}, with an 
imaginary Regge
slope that is related to Newton's constant.

The collisions under study are completely characterized by the changes in the 
trajectories of
the particles and the momentum transfer.
We checked explicitly that the equation of motion gives the longitudinal {\it 
and} the
transverse trajectories. Both effects are generated by the single function $f$.
We were able to calculate the recoil of the particle during the process as well, 
an aspect which
had not been well understood in earlier works.

At the quantum level, the new action gives four non-commuting coordinates, as 
already shown
in \rf{s}, but the commutator is nonlocal. This breaks covariance under general 
world-sheet
reparametrizations. Furthermore, we find that due to the dependence
of the four coordinates on the world-sheet coordinates (that is, the position at 
which
the particles impinge the shockwave), the Heisenberg algebra has to be corrected 
with a term
proportional to $G$, whose form however we could not demonstrate due to 
nonlocality, and is
therefore
still at the speculative stage. At this point the need for new variables became 
evident.
Nevertheless, the proposed correction to the Heisenberg commutator was shown to 
exactly give the
correct momentum transfer that takes place during the collision, and for the 
fact that the
momentum operator $P^\m$ generates translations in the direction $X^\m$. The 
latter is a
consequence of the reduction of the number states of the Hilbert space, since 
momentum and
position operators are identified in a specific way, and the coordinates undergo 
a certain
``merging" due to the
gravitational interaction. We also obtained an algebra for the in and the
out momentum states. This algebra, however, is nonlocal and very nonlinear. Only 
for the total
momentum operator we get a simple algebra, which is the one expected. The
arguments seem to go beyond the
eikonal approximation. In general, all the coordinates have to be treated 
quantum mechanically,
and not just the longitudinal ones. Our results corroborate the arguments by 
Verlinde \rf{erik},
according to which for small initial transverse momenta and for large impact 
parameter, the
eikonal approximation can be used. Indeed, in that case the momentum transfer is 
zero. But for
small impact parameter, the transverse momentum transfer cannot be neglected. 
The longitudinal
momentum transfer is only relevant if the particles collide at nonzero incidence 
angles or if
we go to higher orders in $\ve$.

Dimensionally reducing to $2+1$ dimensions things became much easier. The 
results obtained in
this case confirmed those of the $3+1$-dimensional case, and in particular the 
correction of
the Heisenberg algebra by a term proportional to the momentum. Both results 
agree exactly!
A discrete structure of Hilbert space is found in the lower dimensions.

Another point to study is the Hilbert space structure of the $3+1$-dimensional 
$S$-matrix.
In the presence of interactions, the wavefunctions seem to be modified not 
exactly by a plane
wave factor, but rather by a Chern-Simons factor. The reason is that the
displacements are generated in the directions perpendicular to the motion. Also, 
little can be
said about the finiteness of the number of microstates of the black hole at this 
stage,
especially because, although we do obtain a closed algebra for the different 
operators, there
is still a very nontrivial $\t\s$-dependence. We see two possibilities to say 
more on this
issue. One is to try and find new variables, like in the 2+1 dimensional case, 
to see if one can
get an algebra that generates a discrete spectrum. The other is to take the 
background fields
$B_{\m\n}(X)$ and $\Phi(X)$ seriously, and do a calculation in 2+1 or 
3+1-dimensional gravity
on this background to see what one gets.

Time is still a problem in this $S$-matrix picture. The world-sheet of the
string, corresponding to the horizon, is Euclidean, and the interactions take 
place
instantaneously. The issue of time has arisen several times, in particular when 
discussing
quantisation, suggesting that we should include it in the $S$-matrix from the 
beginning, by
means of step functions which keep track of which particles passed before, 
something similar to
\be
S\smqu-T\int\d^2\t\s\@{h}\,\l[\f{1}{2}\,g_{\m\n}\,\p_iX^\m\,\p^iX^\m
+\l(J^{\sm{out}}_\m\,\th(\lb_{\sm{out}}-\lb_{\sm{in}})+J^{\sm{in}}_\m\,\th(\lb_{
\sm{in}}
-\lb_{\sm{out}})\r)X^\m\r].
\le{actionlambda}
At the end, one would choose a gauge like $\lb_{\sm{in}}=u$ or 
$\lb_{\sm{out}}=v$, depending on
which variable is a good affine parameter along the geodesic. Maybe one
has to integrate the action \eq{actionlambda} over the parameter $\lb$, thus 
obtaining a
three-dimensional world sheet. But all these aspects escape our present 
knowledge.

One would prefer to first have a local theory which one can then covariantly 
generalise. The
approach advocated here, however, although not satisfactory since it contains 
nonlocal
expressions in the intermediate steps, does give results which in the end can be 
expressed
locally and taken as a starting point for an improved theory.
\section*{Acknowledgements}
We would like to thank Gerard 't Hooft and Serge Massar for many interesting 
discussions and
suggestions.
\appendix\section*{Appendix A: Geodesics in the field of a massless particle}
It is interesting to see how one gets the result \eq{l8} from a direct 
calculation of the geodesics in the gravitational field of a particle that 
travels at the speed of light. This is done in reference \rf{gnp85}. The result 
is the shift in the metric mentioned above (see equation \eq{4}), which can be 
used as a starting point for calculating the effect in more general spacetimes, 
as explained in section 1.

Here we briefly review the calculation of the transverse trajectory $y(v)$ of a 
test particle in this field.

One starts with the linearised Schwarzschild metric of a light particle with 
mass $m$. One then calculates the geodesics up to ${\cal O}(m)$ by the 
Euler-Lagrange equations. The result is, for the transverse coordinate $y$,
\be
y(y_0)=y_-(y_0)-2m\,\@{1+\f{\lb^2E^2}{y_0^2}},
\le{A1}
where $\lb$ is the affine parameter along the geodesic, $E$ is the energy of the 
in-going particle and $y_0$ is the transverse distance between both particles as 
they cross each other.

This is the solution obtained when the in-going particle is at rest. However, 
since we are interested in the effect of a massless particle that travels at the 
speed of light, we have to boost the solution by an infinite $\g$-factor and, at 
the same time, take the massless limit. Since the particle is boosted in the 
$z$-direction, the boost does not have any effect on $y$. However, taking the 
massless limit does have an effect. First of all, one has to choose a good 
affine parameter in the boosted frame. It turns out that, since $v$ is 
proportional to the product $m\lb$, outside the plane of the ingoing particle, 
i.e., for $v\not=0$, $\lb$ diverges everywhere in the massless limit. Therefore, 
$v$ is to be taken as the affine parameter rather than $\lb$ outside the 
shockwave. So we write $y$ in \eq{A1} as a function of $v=m\lb\,\f{E}{p}$:
\be
y=y_--2m\,\@{1+\f{v^2p^2}{m^2y_0^2}}=y_--\f{2}{y_0}\,|v|p\,\@{1+m^2}.
\le{A2}
In the massless limit we get
\be
y=y_--\f{2}{y_0}v\,\mbox{sgn}\,(v)\,p
\le{A3}
and
\be
\f{\p y}{\p v}=\f{\p y_-}{\p v}-\f{2}{y_0}\,\mbox{sgn}\,(v)\,p
\le{A4}
which is (in units where $G=1$) the expression \eq{l8}, since in Minkowski space 
the function $f$ is 
logarithmic in the transverse distance $y_0$.
\section*{Appendix B: The induced two-dimensional Ricci tensor}
In this Appendix we calculate the transverse part of the Ricci tensor of the 
metric
\be
\d s^2=2A(u,v)\,\d u\d v+g(u,v)\,h_{ij}(x^i)\d x^i\d x^j,
\le{B0}
defined as
\be
R_{ij}\equiv R^\m_{\;\;i\m j}.
\le{B2}
We use the formula
\be
R^\m_{\;\;i\m j}=\f{1}{\@{-\det g}}\,\p_\m\l(\@{-\det g}\, 
\G^\m_{ij}\r)-\p_i\p_j\l(\log\@{-\det g}\r)-\G^\m_{\n i}\G^\n_{j\m},
\le{B3}
which will slightly simplify the calculation. We notice that one can write
\be
\@{-\det g}&=&Ag\,\@{h}\nn
\G^\a_{ij}&=&-\f{1}{2}\,g^{\a\b}h_{ij}\p_\b g\nn
\G^i_{j\a}&=&\f{1}{2g}\,\dt^i_j\p_\a g\nn
\G^i_{\a j}&=&\f{1}{2g}\,\dt^i_j\p_\a g\nn
\G^\a_{\b i}&=&\G_{i\a\b}=0,
\le{B4}
where the indices $\m,\n$ run from 1 to 4, $\a$ and $\b$ take the values $1$, 
$2$, and $i,j$ take the values $3,4$. Plugging all this in equation \eq{B3}, we 
get:
\be
R_{ij}&=&R^{(2)}_{ij}-\f{1}{2Ag}\,h_{ij}\p_\a\l(Agg^{\a\b}\p_\b 
g\r)-\G^\a_{ki}\G^k_{j\a}-\G^k_{\a i}\G^\a_{jk}\nn
&=&R^{(2)}_{ij}-\f{1}{A}\,h_{ij}\,g_{,\,uv}.
\le{B5}
Here $R^{(2)}_{ij}$ is the two-dimensional Ricci tensor calculated in the metric 
$h_{ij}$.
Now we require $R_{ij}$ to be a solution of the Einstein vacuum equation
\be
R_{ij}&=&0,
\le{B6}
which amounts to \eq{53}.

Next we briefly show how \eq{B6} comes about in the calculation of Dray and 't 
Hooft. Their
method was to start with the metric \eq{B0},
\be
\d\t s^2=2A(u,v)\,\d u\d v+g(u,v)\,h_{ij}(x^i)\d x^i\d x^j.
\le{B1}
This is taken to be a vacuum solution of the Einstein equation ({\it e.g.}, flat 
Minkowski
space). The tilde is used to indicate that this is a {\it different} solution 
from the one we
are interested in, since we look for solutions with a massless particle as a 
source.
One then applies the shift to the $u$-coordinate, that is, one keeps \eq{B1} for 
$v<0$ but
replaces $u$ by $u+\th u$ for $v>0$ (after the particle has fell in). In this 
way one obtains a
different metric,
\be
\d 
s^2=2A(\^u,\^v)\,\d\^v(\d\^u-\dt(v)\d\^v)+g(\^u,\^v)\,h_{ij}(\^x^i)\d\^x^i\d\^x^
j.
\le{C2}
On top applying the shift, we have made the coordinate transformation
\be
\^u&=&u+\th f\nn
\^v&=&v\nn
\^x^i&=&x^i.
\le{C3}
In terms of these coordinates, all the curvature is concentrated at $\^v=0$. 
Outside the plane of the shockwave, we have flat space. 

This metric should be a solution containing a photon located at $\^v=0$, 
$\^x^i=0$. 
Therefore we insert \eq{C3} into Einstein's equation, with a source 
\be
T^{\^u\^u}=4p\,\dt(\^v)\,\dt(\t x).
\le{C3b}
One then gets 
\be
R_{\^{\imath}\^{\jmath}}=R_{ij}^{(2)}-\f{1}{A}\,h_{ij}\l[\^g_{,\,\^u\^v} 
+\^g_{,\,\^u\^u}\,f(\^x^i)\,\dt(\^v)\r],
\le{C4}
and this must be zero, since there is no momentum in the transverse direction. 
Furthermore, since the coordinates $(u,v,x,y)$ satisfied $R_{\m\n}=0$, in 
particular they must 
satisfy $R_{ij}=0$, and using \eq{B6}, we see 
that the first to terms of \eq{C4} amount to this tensor. Therefore they can be 
dropped. We are left with
\be
R_{\^{\imath}\^{\jmath}}=-h_{ij}\f{\^g_{,\,\^u\^u}}{\^A}f(x^i)\,\dt(\^v)=0.
\le{C5}
This, together with the contributions from the other components of the 
Ricci-tensor, gives us the conditions \eq{50}\footnote{In that equation we have 
dropped the carets, since we work only in the shifted coordinate system.}.
\section*{Appendix C: Commutator algebra}
In this appendix we describe the problem mentioned in section 5. Consider the 
operator
identification \eq{eom}, needed for consistent quantisation. Commuting
this equation with $X^\n(\t\s')$, and applying
\eq{80}, gives
\be
&[\L X^\m(\t\s),X^\n(\t\s')]=-\f{i}{2T}\,\e^{\m\n\a\b}\,W_{\a\b}\,\dts.
\le{82}
This algebra is local, but one can easily see that it cannot be integrated to 
give us the
commutator $[X^\m,X^\n]$ if $W^{\m\n}$ explicitly depends on $\t\s$. Naively, 
one would
integrate \eq{82} to give
\be
&[X^\m(\t\s),X^\n(\t\s')]=-\f{i}{2T}\,F^{\m\n}(\t\s')\,f(\t\s-\t\s'),
\le{comm}
where $F^{\m\n}(\t\s)=\e^{\m\n\a\b}\,W_{\a\b}(\t\s)$, but evidently, apart from 
being again
non-local, when $F^{\m\n}$ is really $\t\s'$-dependent, this cannot be right, 
since the right-hand side has to be symmetric with respect to $\t\s,\t\s'$. 
Indeed, starting from the equation of motion for $\L X^\n(\t\s')$ and commuting 
this with $X^\m(\t\s)$ would give us $F^{\m\n}(\t\s)$ instead of 
$F^{\m\n}(\t\s')$ on the right-hand side of \eq{comm}. One can think of 
summing both expressions, or to look for more a general algebra like
\be
&[X^\m(\t\s),X^\n(\t\s')]=-\f{i}{2T}\int\d^2\t\s''\,F^{\m\n}(\t\s'') 
\,g(\t\s,\t\s';\t\s''),
\le{85}
where $g$ is a Green function whose properties can be derived by applying the 
Laplacians $\L$ and $\L'$ to \eq{85}. However, from Fourier analysis one can 
easily see that such a function does not exist.

't Hooft has proposed another expression for the commutator \rf{gnato}, but one 
can easily
check that ---again--- this only works for $F^{\m\n}$ not depending on $\t\s$.

So we were not able to get a consistent algebra for the fields $X^\m$ in the 
general case,
although there may be more general Ansatze than \eq{85}. Therefore, for the 
moment we will work
with the non-local candidate \eq{comm}, which {\it is} valid if $X^\m$ is the 
operator for the
``hard" particles, and $X^\n$ that for the soft ones. In this case, the 
commutator does not need to
be symmetric with respect to $\t\s$ and $\t\s'$, since the $W^{\m\n}$-background 
then refers
to the the ``soft" particle, and the whole picture is consistent. Here we will 
assume that we can
treat both effects (the back-reaction of the in- and of the out-going particles) 
independently.

One can easily see that the effect of the nonlinearities on the right-hand side 
of \eq{comm} is
negligible to first order in $\ve$. Notice that the right-hand side of equation 
\eq{comm} is
already of order $\ve$. Furthermore, from \eq{eom} we learn that, to zeroth 
order in $\ve$, 
$\p_iX^\m$ only depends on $\t\s$ through the Virasoro expansion. So we are 
allowed to use \eq{comm} as long as we put the free oscillations of $X^\m$ 
equal to zero. We are then left with
\be
X^\m(\t\s)=X^\m_0+p^\m_i\s^i+\dt X^\m,
\le{89}
where $p_i$ is some integration constant. In that case, $\p_iX^\m=p^\m_i+{\cal 
O}(\ve)$ and, to
first order in $\ve$, \eq{comm} is indeed valid. The ${\cal O}(\ve)$ terms of 
\eq{89} 
will only have an effect when we investigate the Jacobi identity.

As remarked in \rf{s}, when taking into account the operator nature of 
$F^{\m\n}$, the commutator \eq{comm} together with \eq{80} does not satisfy the 
Jacobi identity:
\be
&[[\L X^\m(\t\s),X^\n(\t\s')],P^\a(\t\s'')] +\mbox{cyclic} =0,
\le{90}
but it is violated by a term proportional to $G$. This suggests that a linear 
term in the
gravitational constant should be added to \eq{80}. The origin of this term is 
the fact that, due 
to the interactions, the momentum of the out-coming particles changes after they 
cross the shockwave, as the coordinate $X^\m$ changes as well. This is explained 
in the section
5.

After some algebra, one can calculate that, in order to satisfy the Jacobi 
identity, \eq{80} has to be modified in the following way:
\be
&[X^\m(\t\s),P^\n(\t\s')]=ig^{\m\n}\,\dts+iA^{\m\n}(\t\s,\t\s'),
\le{heisenb}
with
\be
A^{\m\n}(\t\s,\t\s')\smqu\f{1}{T\@{h}}\,\e^{\m\n\a\b} 
\e^{ij}\p_iX_\a\p_j\dts\int\d^2\t\s''\,P_\b(\t\s'')\,f(\t\s-\t\s'').
\le{Amunu1}
Although, for notational simplicity, we have not written down the labels in or 
out, it is important to keep track of which are the operators corresponding to 
the in or the out-coming particles. $A^{\m\n}$ can be written as
\be
A^{\m\n}(\t\s,\t\s')\smqu&-&\f{1}{T\@{h}}\,\e^{\m\n\a\b}\e^{ij}\p_iX_\a\l(\dts\i
nt\d^2\t\s''\,
P_\b(\t\s'')\,\p_jf(\t\s-\t\s'')\r.\nn
&-&\l.\p_j\dts\int\d^2\t\s''\,P_\b(\t\s'')\,f(\t\s'-\t\s'')\r).
\le{Amunu2}
This expression cannot however be correct. As we argue at the end of section 5 
and
are able to check in section 6, only the first term in this expression can 
contribute. Indeed,
this term, combined with the equations of motion, does give a local algebra; 
furthermore, it
provides the correct momentum transfer and agrees with the 2+1-dimensional 
result when we
compactify one dimension. The reason that we also get the second term is the
awkward $\t\s$-dependence in a nonlocal way, which should be removed by defining 
different
variables. Indeed, equation \eq{90} is very badly defined, since it depends on 
how we let the
Laplacian work on the commutators. So from this one can only get the general 
shape of the
expression \eq{Amunu1}, but not the precise $\t\s,\t\s'$-dependence. The latter 
we fix by the two
arguments already mentioned (correctness of the momentum transfer and comparison 
with the
2+1-dimensional case, where the problem is overcome by defining new variables). 
Therefore, 
we take the first term as an Ansatz, see \eq{AAmunu}. In spite of the fact that 
it contains a
$\dt$-function, the algebra \eq{Amunu} is still nonlocal because of the 
$\t\s$-dependence of
$f$. Nevertheless, after substitution of the equations of motion it will give us 
a local
expression.

We now see two possibilities to solve the problem about how to get the 
commutator \eq{comm}.
The most
promising one is to look for new variables which do not depend on $\t\s$ but 
only on the
boundary of some region of the horizon. On the other hand, the fact that the 
covariant approach
advocated here does not seem to work in general may be a pointer that one has to 
go back to the
original philosophy where one keeps track of what in-going and out-coming 
particles are, and of
which are ``hard" particles and which are ``test" ones, by writing
$X^\m(\t\s)=X^\m_{\sm{in}}(\t\s) +X^\m_{\sm{out}}(\t\s)$, as we have implicitly 
assumed here,
and include theta-functions depending on the parametrisation of the geodesic, 
$\theta(\lb_{\sm{in}} -\lb_{\sm{out}})$ and $\theta(\lb_{\sm{out}} 
-\lb_{\sm{in}})$. This would break the symmetry of \eq{85}, so that the 
conditions on the algebra would be less stringent.

\end{document}